# A mathematical finance approach to the stochastic and intermittent viscosity fluctuations in living cells

**Claude L. Bostoen[1] and Jean-François Berret[2]**

[1]DPI[*], P.O. Box 902, 5600 AX Eindhoven, The Netherlands

[2]Matière et Systèmes Complexes, UMR 7057 CNRS Université Denis Diderot Paris-VII, Bâtiment Condorcet, 10 rue Alice Domon et Léonie Duquet, F-75205 Paris, France

**Abstract**: Here we report on the viscosity of eukaryotic living cells as a function of the time, and on the application of stochastic models to analyze its temporal fluctuations. The viscoelastic properties of NIH/3T3 fibroblastic cells are investigated using an active microrheological technique, where magnetic wires, embedded into cells, are being actuated remotely. The data reveal anomalous transient responses characterized by intermittent phases of slow and fast rotation, revealing significant fluctuations. The time dependent viscosity is analyzed from a time series perspective by computing the autocorrelation functions and the variograms, two functions used to describe stochastic processes in mathematical finance. The resulting analysis gives evidence of a sub-diffusive mean-reverting process characterized by an autoregressive coefficient lower than 1. It also shows the existence of specific cellular times in the ranges 1 - 10 s and 100 - 200 s, not previously disclosed. The shorter time is found being related to the internal relaxation time of the cytoplasm. To our knowledge, this is the first time that similarities are established between the properties of time series describing the intracellular metabolism and statistical results from mathematical finance. The current approach could be exploited to reveal hidden features from biological complex systems, or determine new biomarkers of cellular metabolism.

Corresponding authors:
c.bostoen@polymers.nl
jean-francois.berret@univ-paris-diderot.fr

## 1. Introduction

When observed under an optical microscope, living eukaryotic cells adhering to a substrate appear to be immobile. To observe their displacement, migration or duplication, it is necessary to record their movements by time-lapse imaging. Furthermore, one needs to accelerate the sequence about 100 times to see living cells move, interact with neighbors or perform mitosis. This simple observation suggests that characteristic times for cell motion are of the order of a few minutes or more. However, the above time-lapse sequences alone cannot be used to assess the characteristic times involved in these motions and quantitative experiments are necessary. Over the last two decades, there has been intensive research on the relationship between the mechanical properties of cells and their biological activity.[1-4] Many experiments have been designed to probe the mechanical behavior of the whole cell or its interior, the cytoplasm, including atomic force microscopy (AFM),[5-7] optical and magnetic tweezers,[8-16] and the monitoring of internalized or injected particles.[10,14,17-20] These studies have emphasized the importance of cellular mechanics in some basic functions such as proliferation, differentiation, adhesion and migration. Cell biomechanics is also suspected of being involved in the spread and invasion of metastatic cells leading to the formation of distant secondary tumors.[21,22] Despite the large amount of work done in the field of cellular biomechanics, no definitive consensus has been reached regarding the rheological model for the cytoplasm, or the times characterizing cell motions.[2,4,23]

---

[*]Independent research

To probe long-term dynamics in rheology, it is imperative to explore the low frequency region (down to $10^{-3}$ rad s$^{-1}$) of the complex modulus $G^*(\omega) = G'(\omega) + iG''(\omega)$, where $G'(\omega)$ and $G''(\omega)$ denote the storage and loss moduli, respectively.[24] For relaxation times greater than 100 s, other approaches such as temperature-frequency superposition in the polymer melts or stress relaxation experiments are suitable. At the cell level, however, low frequency microrheology experiments are scarce and most $G^*(\omega)$-measurements reported to date, either from passive or active microrheology have been obtained at an angular frequency above 0.1 rad s$^{-1}$.[8-12,17,18,25] Recent approaches using a high-speed AFM spectroscopy even make it possible to reach the $10^6$ rad s$^{-1}$ range.[26] As a result, living cells from different cell lines have been found to be characterized by $G'(\omega)$ larger than $G''(\omega)$ and for this reason described as soft elastic materials. With respect to slow metabolic processes that may affect slow intracellular dynamics, Wottawah *et al*. have applied step stress deformation on an entire cell (an equivalent of creep rheology) and found overall relaxations of the order of seconds, which they attributed to the dissociation of the actin network.[27] Also related to actin, dynamical processes involving periodic contractions of the lamellipodia or actin waves travelling during migration have been found on a time scale of a few tens of seconds.[28,29] More recently, an AFM investigation has shown evidence of water transport in the cytoplasm on a time scale of seconds, consistent with a poroelastic description of the cytoplasm.[7] Concerning elastic modulus measurements, force-indentation performed on live human bronchial cells has revealed a cyclic activity of the cytoskeleton with a period of 200 s, that was ascribed to collective action of myosin motor proteins.[15,16] The above results are however insufficient to conclude on the long-term internal dynamics of living cells.

Following the pioneering work of Crick and Hughes on the physical properties of the cytoplasm,[30] we have proposed a technique based on the remote actuation of calibrated micron sized wires to probe intracellular dynamics. In this experiment, a magnetic field rotates the internalized wire in the cytoplasm in a propeller-like motion at angular frequency $10^{-3}$ to 10 rads$^{-1}$.[31,32] At frequencies below $10^{-2}$ rad s$^{-1}$, a purely viscous behavior was found in three cell lines, murine NIH/3T3 fibroblasts, HeLa cervical cancer cells and A549 lung carcinoma epithelial cells.[32] Above 1 rad s$^{-1}$, the wires exhibited an elastic response in agreement with the aforementioned work.[8,11,12,26] The cell lines put under scrutiny here were shown to have shear viscosities in the range 10 - 100 Pa s, and internal relaxation times of 3 to 30 s. In the sense of rheology and in the sequel of the paper, the internal relaxation time is obtained from the ratio of the static viscosity by the elastic modulus.[24] With the above technique, the viscosity data were averaged over the duration of the experiments (typically a few minutes), and compared to the known constitutive equations, e.g. that of the Generalized Maxwell model.[32] In the present work, we reconsider the experimental viscosity measurements, *i.e.* we focus on its time dependent nature. In a first step, the time dependent viscosity of NIH/3T3 mouse fibroblasts is derived and analyzed from a time series perspective. Subsequently, we apply to the data the statistical tools used in mathematical finance and econophysics[33,34] to describe stochastic processes present in the financial markets.[35-37] Following the approach by Bouchaud and Potters,[33] we calculate for each angular frequency the autocorrelation function and the variogram, using the time series of the logarithm of the viscosity. These functions offer the advantage of studying possible random walk, autoregressive and mean reverting behaviors. In mathematical finance, the cumulative distribution of prices or interest rates has been studied extensively,[33] revealing the existence of so-called fat tails which were well described by a Student's t-distribution. A similar analysis of the time series of the cytoplasm viscosities is presented here.

## 2. Methods
### 2.1. Magnetic Rheometer
The magnetic wire micro-rheology technique has been described in previous accounts.[38-40] Fibroblast cells were incubated with magnetic wires[31] at a 1:1 ratio and then sealed in a Gene Frame® (Abgene/Advanced Biotech, dimensions 10×10×0.25 mm$^3$). The glass slide was introduced into a homemade device generating a rotational magnetic field, thanks to two pairs of coils (23 ohms) working with a 90°-phase shift. An electronic set-up allowed measurements in the frequency range $\omega = 10^{-3} - 10^2$ rad s$^{-1}$ and at magnetic fields $\mu_0 H = 0 - 20$ mTesla. The microrheology protocol used is based on the Magnetic Rotational Spectroscopy technique.[32,40-42] Phase-contrast and bright field images were acquired on an IX73 inverted microscope (Olympus) equipped with 20× and 100× objectives. An EXi Blue camera (QImaging) and Metaview software (Universal Imaging Inc.) were used as acquisition system. For each condition of magnetic field and angular frequency, a movie was recorded for a period of time of 300 s or longer and then treated using the ImageJ software (https://imagej.nih.gov/ij/). For the wire calibration, magnetic wire actuation was performed on a series of water-glycerol mixtures at T = 25 °C and glycerol concentrations of 49.8%, 81.0%, 84.5% and 89%, leading to a susceptibility anisotropy coefficient $\Delta\chi = 2.3 \pm 0.7$. Phase-contrast and bright field images were acquired on an IX73 inverted microscope (Olympus) equipped with 20× and 100× objectives. An EXi Blue camera (QImaging) and Metaview software (Universal Imaging Inc.) were used as acquisition system. The cell culture conditions were identical to those described in Refs.[32,43]

Wire rotation may be interpreted in terms of several mechanical models, assuming that the surrounding sample exhibits either purely viscous, viscoelastic or purely elastic responses. The validity of approach and data analysis was confirmed in experiments performed on wormlike micellar solutions and polysaccharide gels of known rheology. The wire-based microrheology has shown that the data are consistent with constitutive equations derived from the linear response theory. Moreover, the values of the viscosity and elastic moduli obtained on these model fluids were con-firmed by frequency sweep tests made with cone-and-plate rheology in the linear regime of deformation.[38,40]

**2.2. Data analysis**

The data treatment and analysis were done with RStudio, using the open source data analysis software R.[44] We have used the autocorrelation and variogram statistical functions to analyze intracellular viscosity time series. For the fitting, the Generalized Random Walk model, with and without the Hole effect and the Ornstein-Uhlenbeck Process model were applied. This combination allowed a cross-determination of the different parameters introduced to describe a sub-diffusive, mean-reverting process and periodical dynamics in cytoplasm. The models are described in the main text in the section *Generalized Random Walk data analysis*.

# 3. Results and Discussion
## 3.1. Cellular Microrheology

In this section, the tools and methods used to perform the active microrheolgy are presented and it is shown how to translate the wire motion into the time dependent viscosity are presented. Fig. 1a displays a phase contrast optical microscopy of magnetic wires subjected to a field $\mu_0 H = 10$ mTesla. The wires have lengths comprised between 5 and 50 µm and an average diameter of 1 µm. The alignment of the wires with an external magnetic field (arrow) demonstrates their magnetic properties.[45] To make the wire lengths compatible with the dimensions of the cells, the suspension is sonicated for a few minutes, resulting in a decrease in the median length to 2.4 µm and in a dispersity of 0.35 (**Supplementary Information S1**).[32] When fibroblasts are exposed to wires in Petri dishes at a 1:1 wire-to-cell ratio, we observe that they sediment on the fibroblast layer due to their own weight, and over time spontaneously enter into the cytoplasm. The toxicity measurements show that at the doses used, the cells retain excellent viability.[43] Fig. 1b displays a group of 7 cells observed in phase contrast microscopy with wires inside cells (yellow arrows) and wires which are lying on the bottom wall of the Petri dish (green arrow). The internalization of magnetic wires inside cells is illustrated in **Supplementary Movie#1** and in **Supplementary Information S2**.

Once internalized, the wires are subjected to a rotational magnetic field with an angular frequency $\omega$ between $10^{-3}$ and 10 rad s$^{-1}$, and their orientation is monitored using time-lapse optical microscopy. In Fig. 1c the orientation angle $\theta(t)$ of a 2.8 µm magnetic wire is plotted as a function of time at the frequency of 0.015 rad s$^{-1}$. There, $\theta(t)$ increases linearly at the same angular speed as the field, indicating synchronous rotation. At higher frequency ($\omega = 0.65$, 1.3 and 5.3 rad s$^{-1}$), $\theta(t)$ displays back-and-forth oscillations characteristic of an asynchronous regime.[46-49] As shown in previous accounts,[32,38-40] the transition between synchronous and asynchronous regimes at the critical angular frequency $\omega_C$ allows to estimate the cytoplasm viscosity, using:

$$\omega_C = 3\mu_0 \Delta\chi H^2 / 8\eta L^{*2} \tag{1}$$

where $L^* = L/D\sqrt{g(L/D)}$ and $g(x) = ln(x) - 0.662 + 0.917/x - 0.050/x^2$.[38,39]

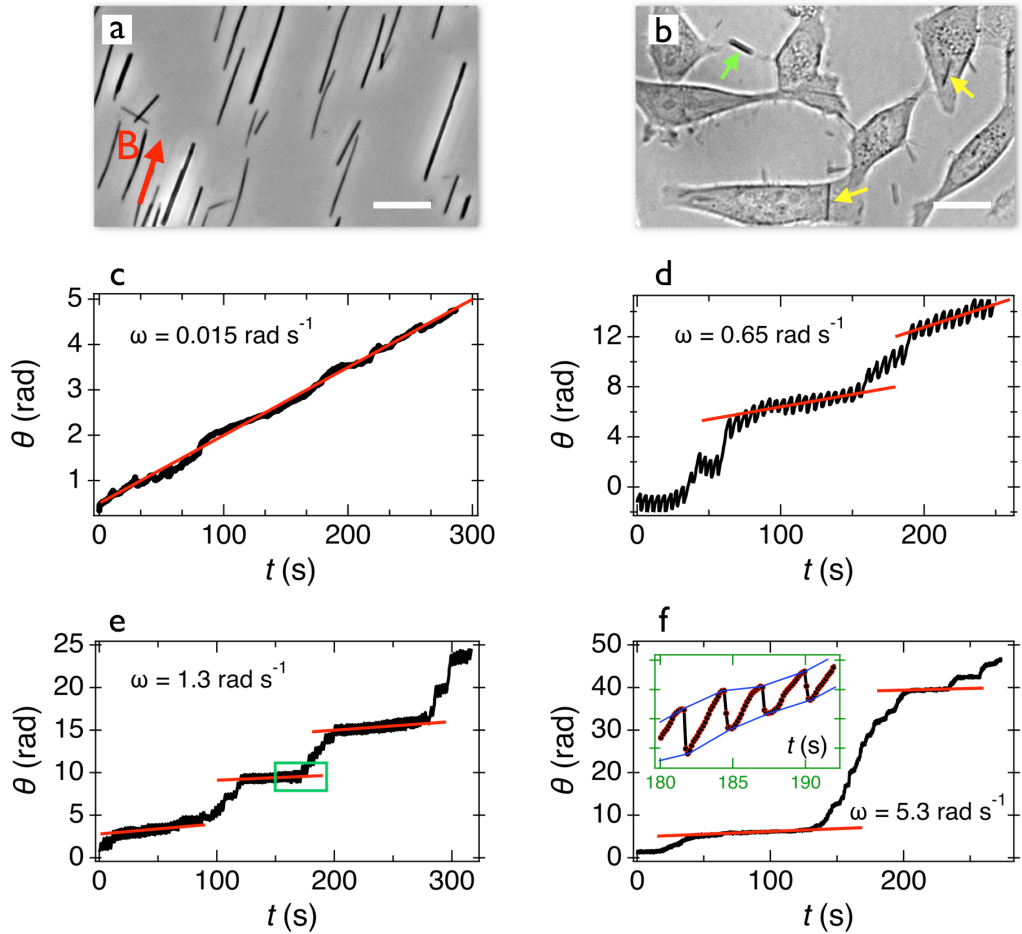

*Figure 1*: **a)** *Phase contrast optical microscopy image of magnetic wires under a static magnetic field. The red arrow is the direction of the magnetic field $\mu_0 H$ = 10 mTesla (magnification ×20, the bar is 10 μm).* **b)** *Optical microscopy images of NIH/3T3 fibroblasts after their incubation with magnetic wires (magnification ×20, the bar is 10 μm). The yellow arrows show wires inside cells, whereas the green arrow points to a wire that sits on the bottom wall of the Petri dish.* ***Supplementary Movie#1*** *illustrates the incubation and internalization kinetic process.* **c)** *Orientation angle $\theta(t)$ as a function the time for a 2.8 μm wire submitted to a magnetic field $\mu_0 H$ = 14 mT at the frequency $\omega$ = 0.015 rad s$^{-1}$. The straight line in red determines the average rotation frequency $\Omega$, which in this case is the applied frequency. The wire is here synced with the field.* **d,e,f)** *Same as in Fig. 1c for angular frequencies $\omega$ = 0.65, 1.3 and 5.3 rad s$^{-1}$. The inset in Fig 1f emphasizes the wire oscillations characteristic of the asynchronous regime, as well as lines determining the instantaneous rotation frequency $\Omega(t)$.*

In the example of Figs. 1c-f, $\omega_C$ is estimated at 0.13 rad s$^{-1}$ and corresponds to a static viscosity of 28 ± 6 Pa s. In the synchronous regime, the transient response exhibits deviations from the linear behavior (Fig. 1c) and in the asynchronous regime it displays an alternation of slow and fast rotation. This is described in the following as an intermittent behavior and related to underlying viscosity fluctuations.[50] The lines in red displayed in the figures describe time domains over which the average rotation frequency denoted $\Omega(\omega) = \langle d\theta(t)/dt \rangle_t$ is constant. These periods last from about 10 to 100 seconds. Between these intervals, the wire accelerates and the average rotation velocity is significantly higher. In our previous work,[32] it was shown that the $\Omega(\omega)$-values shown by the straight lines in Fig. 1 obey the constitutive model predictions obtained for Maxwell-type viscoelastic fluids, namely:[32,38]

$$\begin{aligned}\omega \leq \omega_C \quad & \Omega(\omega) = \omega \\ \omega \geq \omega_C \quad & \Omega(\omega) = \omega - \sqrt{\omega^2 - \omega_C^2}\end{aligned} \quad (2)$$

Fig. 2 illustrates the agreement between experimental data and Eq. 2 for the 2.8 μm wire. Eq. 2 shows that the function $\Omega(\omega)$ depends on the viscosity only in the asynchronous regime (through its relation with $\omega_C$) and not in the synchronous regime. In the following, we exploit this property to examine the wire transient response in the regime $\omega \geq \omega_C$, as well as to interpret the intermittent cell response in terms of time-dependent viscosity. At this point, it is important to recall that the intermittency phenomenon found here is not observed in non-living viscoelastic fluids.[39,46-49]

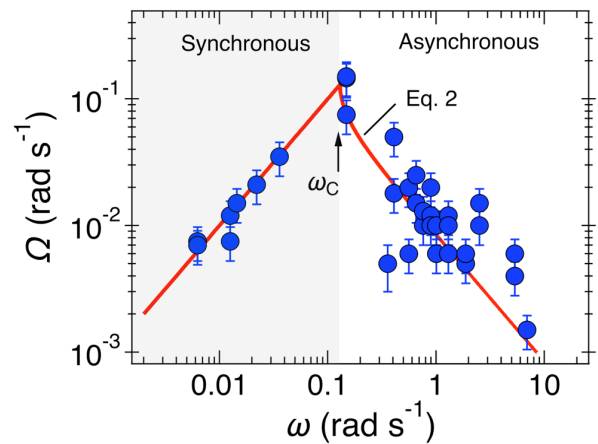

**Figure 2:** *Wire average rotation velocity $\Omega(\omega)$ measured for a 2.8 μm magnetic wire in a NIH/3T3 fibroblast cell as a function of the actuating frequency ω. The continuous line in red results from best fit calculations using Eq. 2 and a critical frequency $\omega_C$ = 0.13 rad s$^{-1}$. The cusp at $\omega_C$ indicates the transition between the synchronous and asynchronous regimes.*

To retrieve the time-dependent viscosity, the $\theta(t)$-traces are analyzed by computing the instantaneous angular velocity $\Omega(t) = d\theta(t)/dt$ at the time scale of the oscillations, that is every $\pi/\omega$ seconds. We thereby assume that Eq. 2, which describes the stationary regime, is also valid at the time scale of the oscillations, leading to $\Omega(t) = \omega - \sqrt{\omega^2 - \omega_C(t)^2}$. In the previous expression, the angular field frequency $\omega$ is fixed and the fluctuations in the wire response are solely due to viscosity changes, which in turn induce temporal changes in $\omega_C(t)$. With these assumptions, $\omega_C(t)$ and $\eta(t)$ are estimated from the instantaneous derivative $d\theta(t)/dt$ and from Eq. 1. To increase the data sampling, this instantaneous slope is calculated from the minima and maxima of the oscillations, as indicated by the blue segments in the inset of Fig. 1f. Fig. 3 provides examples of $\omega_C(t)$ and $\eta(t)$ time series for a 2.8 µm magnetic wire at angular frequencies 0.65, 1.3 and 5.3 rad s$^{-1}$. The data are computed from the $\theta(t)$-traces presented in Fig. 1. Plotted in semilogarithmic scales, both quantities exhibit strong temporal fluctuations around a mean value, indicated by horizontal straight lines. For the viscosity, $\bar{\eta}$ is evaluated at 21.7, 16.0 and 4.2 Pa s, respectively. These fluctuations are associated with standard deviations of the order of the mean. Another important observation from Figs. 3b, 3d and 3f is that time intervals with large and small fluctuations alternate in an apparently random fashion, a feature referred to volatility clustering and being studied in quantitative finance.[33,35] Additional critical frequency and viscosity time series from the same wire are available in **Supplementary Information S3**.

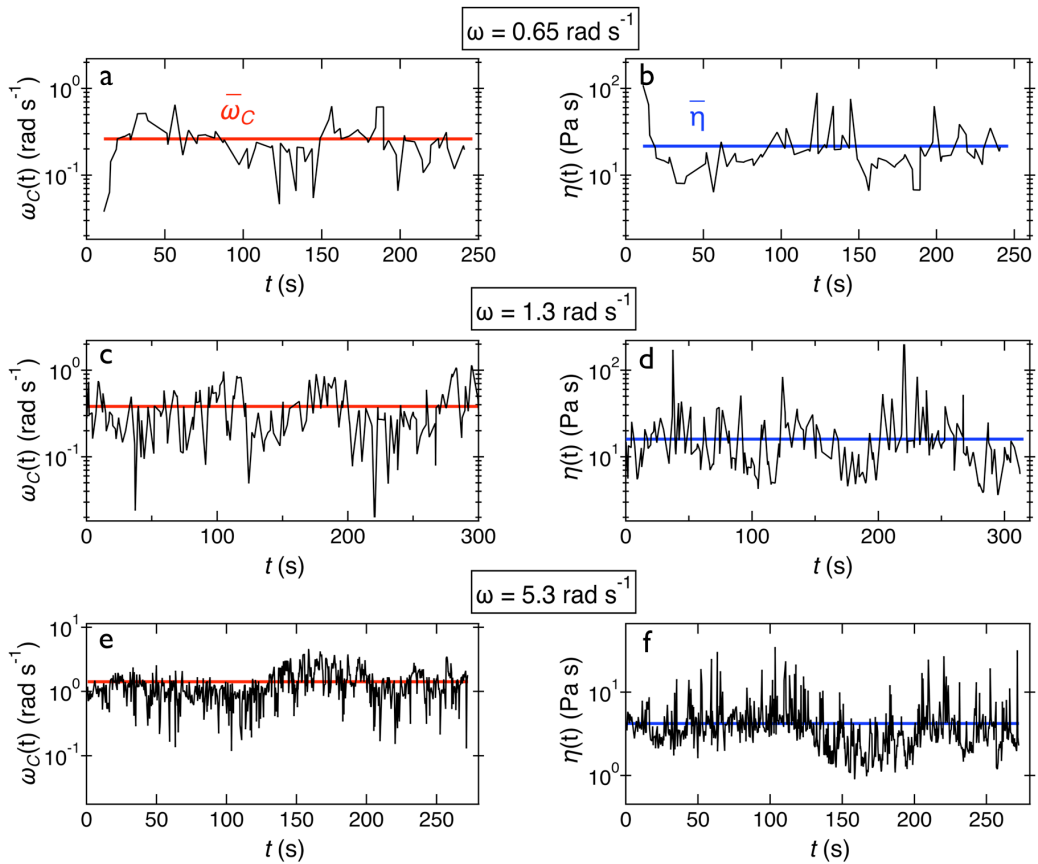

**Figure 3**: *Time dependences of the critical frequency $\omega_C(t)$ (a,c,e) and of the shear viscosity $\eta(t)$ (b,d,f) obtained at the angular frequencies $\omega$ = 0.65, 1.3 and 5.3 rad s$^{-1}$, as indicated. The horizontal lines in the $\omega_C(t)$ and $\eta(t)$-plots are the time series average values. Complementary data obtained at lower and larger frequencies are shown in* **Supplementary Information S3**.

Fig. 4 shows the cytoplasm viscosity *versus* frequency for three different wires and indicates a thinning behavior, an outcome that is reminiscent of a Maxwell fluid studied at high frequency where $G'(\omega) > G''(\omega)$.[24] To our knowledge, this is the first experimental study on cells that finds a viscosity decrease as a function of the actuation frequency. In these instances, the low frequency $\eta$-values determined from $\omega_C$ (*via* Eq. 1) and indicated by arrows are consistent with viscosity data calculated from time averages. In conclusion, the active wire microrheology technique applied to fibroblast cells reveals a generic behavior for the time-resolved viscosity, characterized on a short time scale (seconds) by large fluctuations and on a longer time scale (minutes) by a well determined average value.

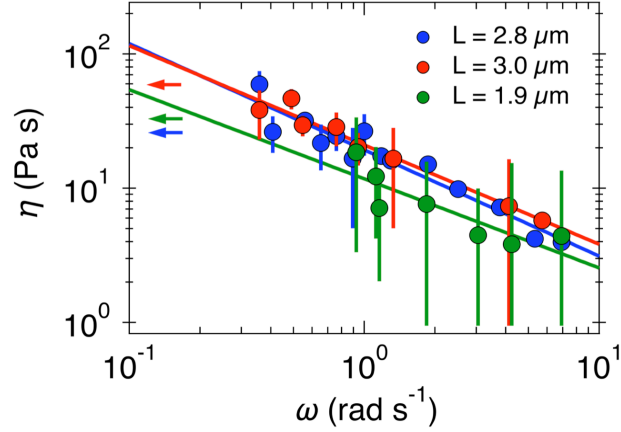

**Figure 4**: *Time average cytoplasm viscosity $\eta(\omega)$ as a function of the angular frequency for magnetic wires of different lengths. The error bars correspond to the standard deviation.*

### 3.2 – Generalized Random Walk data analysis

In this section, the procedure for the cytoplasm viscosity data treatment and interpretation is outlined. For each frequency, a plot of the viscosity as a function of time is made on a linear scale to have a view of the high viscosity values, called outliers and to inspect the data for long-term trends and periodic features. The specifications for eliminating the outliers are provided in the **Supplementary Information S4**. The first step of processing involves removal of the outliers, de-trending the time series and making of the viscosity data equidistant. In the following, we take an approach pointed out in quantitative finance[35-37] and in econophysics,[33,34] which consists in defining statistical quantities using the logarithm of the viscosity $\ln \eta(k)$, where $k$ is the number in a time series with a total of $k_{Max}$ measurements ($1 < k < k_{Max}$). The autocorrelation function $C(l)$ of the logarithm of the viscosities is obtained at each lag $l$ using:[33]

$$C(l) = < \ln \eta(l+k) \ln \eta(k) > - < \{\ln \eta(k)\}^2 > \tag{3}$$

Referring to the function in R, we will use the abbreviation ACF, which is $C(l)$ divided by the variance of the time series.[44] In this approach the time between two consecutive lags decreases with increasing probing frequency as $k_{Max}/\omega$, provided measurement times are equidistant. Figs. 5a1, 5b1 and 5c1 show the ACFs for the cleaned viscosity time series of Fig. 3, respectively for $\omega$ = 0.65, 1.3 and 5.3 rad s$^{-1}$. They exhibit a rapid initial decay at low $l$, followed by an oscillatory behavior typical for periodic correlations. In econophysics, it is advised to calculate the experimental variograms $\gamma(l)$ for times series where the mean is *a priori* not known:[33]

$$\gamma(l) = < \{\ln \eta(l+k) - \ln \eta(k)\}^2 > \tag{4}$$

The $\gamma(l)$'s calculated at the same three frequencies are displayed in the second column of Fig. 5. There, the main features are a step-like increase observed at low lags, followed by a persistent periodic oscillation extending over the entire lag range. Interestingly, the ACF- and $\gamma(l)$-data reveal the existence of two well separated lag scales, noted $\tau_l$ and $T_l$ and indicated by arrows: the first one is associated with the initial exponential decay of the ACF and with the rise of the variogram, whereas the second is linked to the period of the oscillations. In Fig. 5b1 and 5b2, the period $T_l$ corresponds to about 60 lags. The time difference per lag being 1.67 seconds at this frequency, one gets a time period $T$ of about 100 s. For the interpretation of the variograms, we used the approach outlined by Bouchaud and Potters in their work on financial markets, one of the examples of a complex systems.[33] To this aim, a time series using an autoregressive scheme for the logarithmic viscosities is constructed using the expression:

$$\delta(k+1) = \alpha\, \delta(k) + \xi_k, \text{with } \alpha \leq 1 \text{ and } \delta(k) = \ln \eta(k) - <\eta> \quad (5)$$

With $\alpha = 0$, the series becomes the identical independently distributed $\xi_k$ with a variance $\sigma^2$, whilst the case $\alpha = 1$ describes a random walk. The autoregressive equation for $\delta(k)$ shown in Eq. 5 is equivalent to that outlined in Bouchaud et al.[33] and leads to the definition of the variogram in Eq. 4. Here we will use the notion of Generalized Random Walk (GRW) for which the $\alpha$-values are between 0 and 1 and describe the case of sub-diffusive dynamics. The above Eqs. 4 and 5 lead to a recursion relation of the form:

$$\gamma_{GRW}(l) = 2\, \sigma^2 \frac{1 - \alpha^l}{1 - \alpha^2} \quad (6)$$

To extend the predictions of GRW model, we consider a complementary approach based on the Ornstein-Ühlenbeck Process (OUP) model. The OUP model is the continuous time limit of the GRW model and is obtained for $\alpha$ close to unity. Putting $\alpha = 1 - \epsilon$ with $\epsilon \ll 1$ in Eq. 6 leads to:

$$\gamma_{OUP}(l) = \frac{\sigma^2}{\epsilon}(1 - e^{-\epsilon l}) \quad (7)$$

For $l \ll 1/\epsilon$, the correlation time, one finds $\gamma_{OUP}(l) = \sigma^2 l$, the random walk limit. In the opposite case, when $l \gg 1/\epsilon$, the variogram saturates at $\sigma^2/\epsilon$. Using the experimental variogram values of the first two lags, and the ratio of $\tilde{\gamma}_{GRW,OUP} = \gamma(2)/\gamma(1)$, the key parameters $\alpha$ and $\epsilon$ for the GRW and the OUP model can be determined. Note that the $\tilde{\gamma}_{GRW,OUP}$ are numbers independent of the variance $\sigma^2$ of the identical independently distributed $\xi(k)$. For the Generalized Random Walk, this results in:

$$\alpha = \tilde{\gamma}_{GRW} - 1, \qquad \sigma^2 = \frac{\gamma(2)}{2} \quad (8)$$

In the case of the Ornstein-Uhlenbeck Process, one obtains:

$$\epsilon = ln\left(\frac{1}{\tilde{\gamma}_{OUP} - 1}\right), \qquad \sigma^2 = \epsilon \frac{\gamma(1)}{2 - \tilde{\gamma}_{OUP}} \qquad (9)$$

The correlation time, $\tau_l = 1/\epsilon$, can be determined using:

$$\tau_l = 1/ln\left(\frac{1}{\tilde{\gamma}_{OUP} - 1}\right) \qquad (10)$$

To account for the oscillations found in Fig. 5, Eq. 6 is modified by adding an oscillating function of the lag as provided by the Hole effect model:[51]

$$\gamma_{GRW-HE}(l) = 2\sigma^2 \frac{1 - \alpha^l}{1 - \alpha^2} + A_T\left\{1 - \cos\left(\frac{2\pi}{T_l}l\right)\right\} \qquad (11)$$

where $A_T$ is the amplitude and $T_l$ the period describing the Hole effect model. This so-called additive nested variogram was used in geology to determine periodic profiles of alternating strands in measured ore concentrates.[51] In econophysics, Eq. 6 describes and predicts the volatility clustering and mean reverting behaviors observed in financial markets.[33,35-37] Note that the de-trending is necessary for obtaining a stationary time series and the application of the models outlined previously. Eqs. 6 and 11s are then used to fit the experimental variograms of Fig. 5, with the parameters $T_l, \alpha, \sigma^2$ and $A_T$ as adjustable. We thereby evaluate the adequacy of the previous GRW and GRW-HE models with the data by examining the time series at $\omega$ = 0.65, 1.3 and 5.3 rad s$^{-1}$ displayed in Figs. 1, 2 and 3. In Figs. 5a2, 5b2 and 5c2, the continuous lines in blue and in red display least-squares calculations using the functions $\gamma_{GRW}(l)$ and $\gamma_{GRW-HE}(l)$, respectively. Eq. 11 shows an excellent agreement with the data obtained at the three frequencies. In the above examples, the $\alpha$'s are found at 0.58, 0.87 and 0.61, indicating that the viscosity fluctuations are linked to a sub-diffusive process. For sake of completeness, we also included in **Supplementary information S5** the list of parameters that can be retrieved from the Generalized Random Walk and Ornstein-Uhlenbeck Process (OUP) models from fitting ACF and variogram functions. To demonstrate the uniqueness of the time series, the data were randomized and the variogram was again computed. This time, the $\gamma(l)$ does not show any in initial increase or periodic features, but a flat line in agreement with random noise predictions (Fig. 6).[33] In the next section, it is shown that the application of the two models enable a cross-determination of the fitting parameters and a final validation of these results. Last but not least, we apply the nested variogram model of Eq. 11 to measurements made in a different context, that of optical tweezers measuring the time dependent intracellular elastic modulus of Hela cells.[12,52] The elastic modulus time series was found to display a sub-diffusive mean-revering behavior and a persistent periodic oscillation over the entire lag range (**Supplementary information S6**). These results, like those obtained on the cytoplasm viscosity, show a great similarity in their stochastic behaviors, and an excellent agreement with the GRW models.

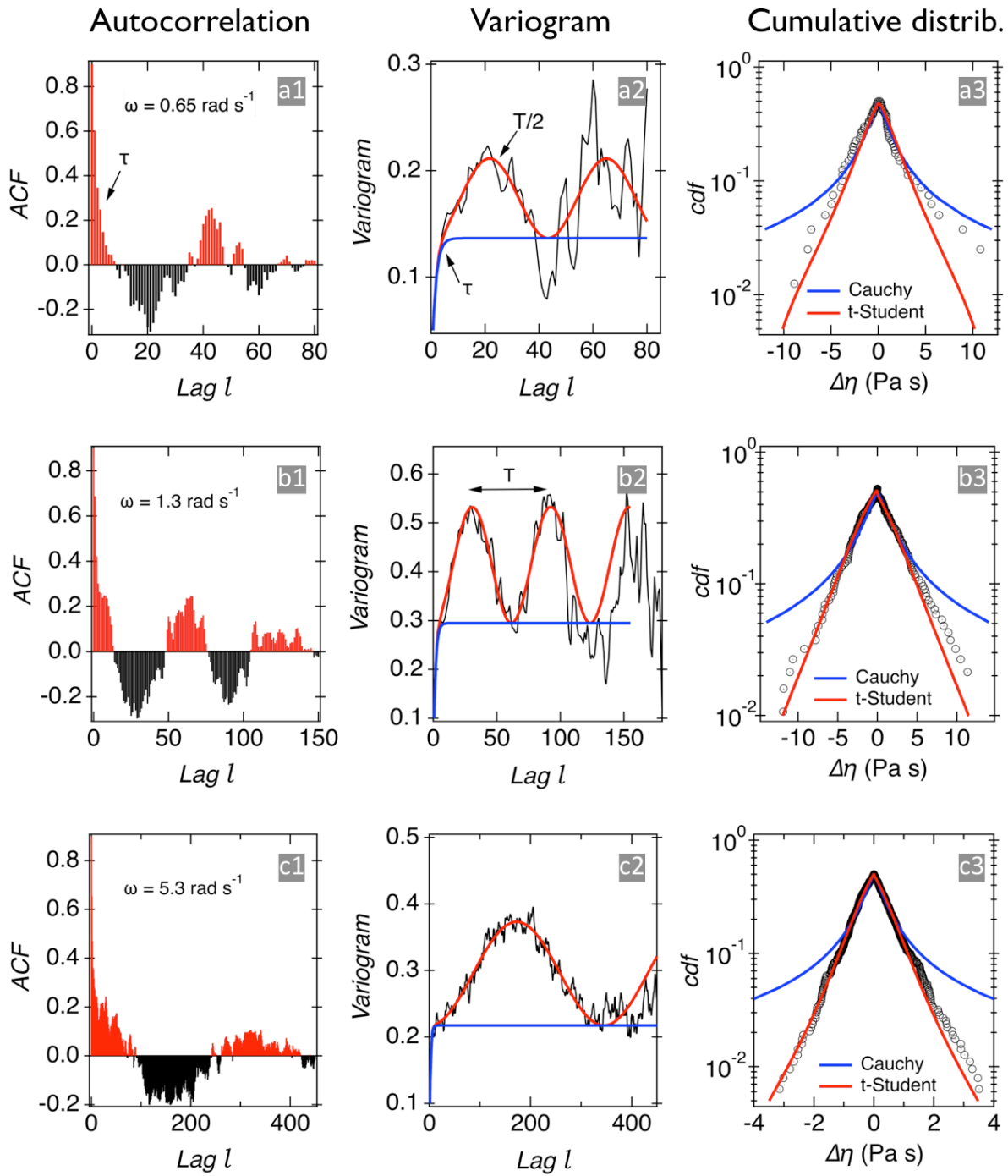

***Figure 5***: ***a1-c1)*** *Auto-correlation functions (ACFs) of the viscosity time series computed from Eq. 3 for three the angular frequencies, ω = 0.65, 1.3 and 5.3 rad s$^{-1}$. The data were obtained after removal of the outliers at the high end, the de-trending and a treatment making data equidistant.* ***a2-c2)*** *Variograms computed from Eq. 4 for the same η(t) time series. The variograms are fitted with the Generalized Random Walk model (Eq. 6, continuous line in blue) and with the GRW model modified by a periodic function (Hole effect model,$^{51}$ Eq. 11, continuous line in red).* ***a3-c3)*** *Experimental cumulative distribution function (cdf) of the viscosity variations Δη at ω = 0.65, 1.3 and 5.3 rad s$^{-1}$. The Cauchy and Student's t-distributions are shown for comparison. The Gaussian functions are displayed in **S7**.*

In a second approach, we have plotted in Figs. 5a3, 5b3 and 5c3 the cumulative distribution function (cdf) of the viscosity differences (*i.e.* for $l = 1$) determined at each frequency. Results indicate that the tails of these distributions are broader than Gaussians, but less broad than the Cauchy distributions displayed in the figures. The Gaussian fits have been evaluated and for sake of clarity they are shown separately in **Supplementary Information S7**. Finally, we found that the observed exponential decrease is best accounted for using a Student's t-distribution. For both variograms and cumulative distribution functions, it is found that the intracellular viscosity fluctuations exhibit strong similarities with those of financial markets, including assets, interest or exchange rates reported in Ref.[33,34] To our knowledge, this is the first time that similarities of this kind have been established between the properties of time series describing the intracellular metabolism and statistical results obtained from econophysics studies.

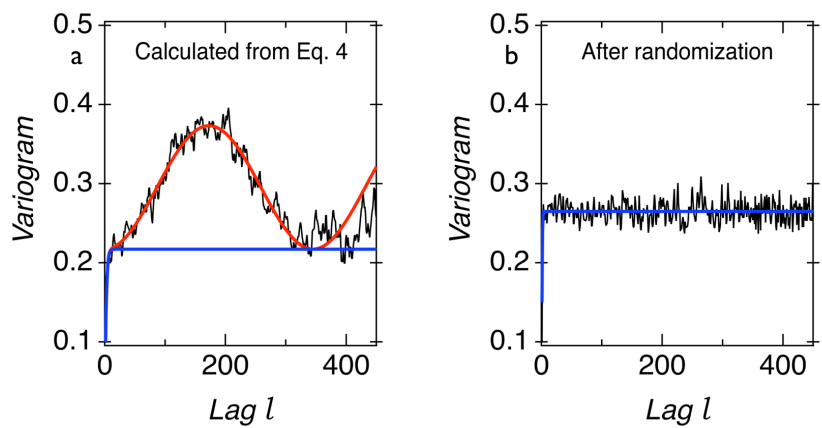

**Figure 6**: *Comparison between variograms obtained at 5.3 rad s$^{-1}$ using viscosity time series as received from **a)** experiments (see Fig. 5c2) and **b)** after data randomization. The initial increase as well as the oscillations do not show up after the randomized viscosity profile.*

### 3.3. Sub-diffusive dynamics, long term correlations and large viscosity fluctuations scaling

The data treatment outlined above has been executed for intracellular viscosity measurements recorded at different angular frequencies in the non-synchronous regime (0.1 – 10 rad s$^{-1}$) and with wires of different lengths (1.9 μm – 3 μm). The initial decay of the autocorrelation function was fitted with a single exponential function, whereas non-linear least squares adjustments were carried out on variograms with the Generalized Random Walk model. Here we discuss the outcomes obtained for the 2.8 μm wire in Figs. 1-3 and refer to **Supplementary Information S8 and S9** for extended results. Fig. 7a displays the frequency dependence of the decay time $\tau$ derived from the ACFs. It is found in the range 0.5 – 10 s to decrease as $\tau(\omega) \sim \omega^{-0.87}$, shown as a straight line in the figure. Also included in the figure for comparison are the outcomes of the ratio $\eta(\omega)/G_0$, where the $\eta(\omega)$ denotes the shear viscosity taken from Fig. 4 and $G_0$ the cytoplasm modulus.[32] $G_0$ was derived from the measurements of the oscillation amplitude $\theta_B(\omega)$ recently reported for this cell line and using the expression $\lim_{\omega \to \infty} \theta_B(\omega) = 3\mu_0 \Delta \chi H^2 / 4 G_0 L^{*2}$ [32,40] At low frequency, $\eta(\omega)/G_0$ tends towards the intrinsic rheological time of the intracellular medium. It is found moreover that $\eta(\omega)/G_0$ matches precisely the autocorrelation time $\tau(\omega)$ over the whole frequency range. This outcome suggests that the time over which the values of the local viscosity are correlated is of the same order than the cell intrinsic rheological time. In contrast, the four other adjustable parameters, *i.e.* the period $T$, the sub-diffusive coefficient $\alpha$, the variance $\sigma^2$ and the Hole effect amplitude $A_T$ show a quasi-independence of the measuring angular frequency (Fig. 7b-7e). The long-time period affecting the oscillations observed in the ACF and in the variograms is found at $T = 124 \pm 39$ s, whereas the $\alpha$ coefficient is $0.57 \pm 0.17$. As $\alpha < 1$, this implies that the dynamics of the time series based on the logarithm of the viscosity is consistent with a sub-diffusive regime. Sub-diffusive diffusion properties have been reported in equilibrium and in non-equilibrium complex systems.[12,52-56] In Fig. 7e, the amplitude $A_T$ is compared to $2\sigma^2/(1-\alpha^2)$, which can be viewed as a measure of the random noise contribution. For some frequencies, the random fluctuations were significant and disturb the periodic processes, leading to less precise determination on the oscillations. The constancy of GRW parameters suggests that the dynamics underlying the cellular activity remain frequency independent. In **Supplementary Information S9**, the fitting parameters obtained using the GRW and the OUP model with wires of length 2.8, 3.0 and 1.9 μm are compared and found in a good agreement. Table I recapitulates the major findings retrieved from the fittings, providing values of the parameters $T$, $\alpha$, $\sigma^2$ and $A_T$ for the three conditions and averaged over all angular frequencies. Also included are $\tau^*$, the ACF relaxation time at $\omega = 1$ rad s$^{-1}$ and $\delta$ the exponent of the scaling law $\tau(\omega) \sim \omega^\delta$ (see **S11** for details).

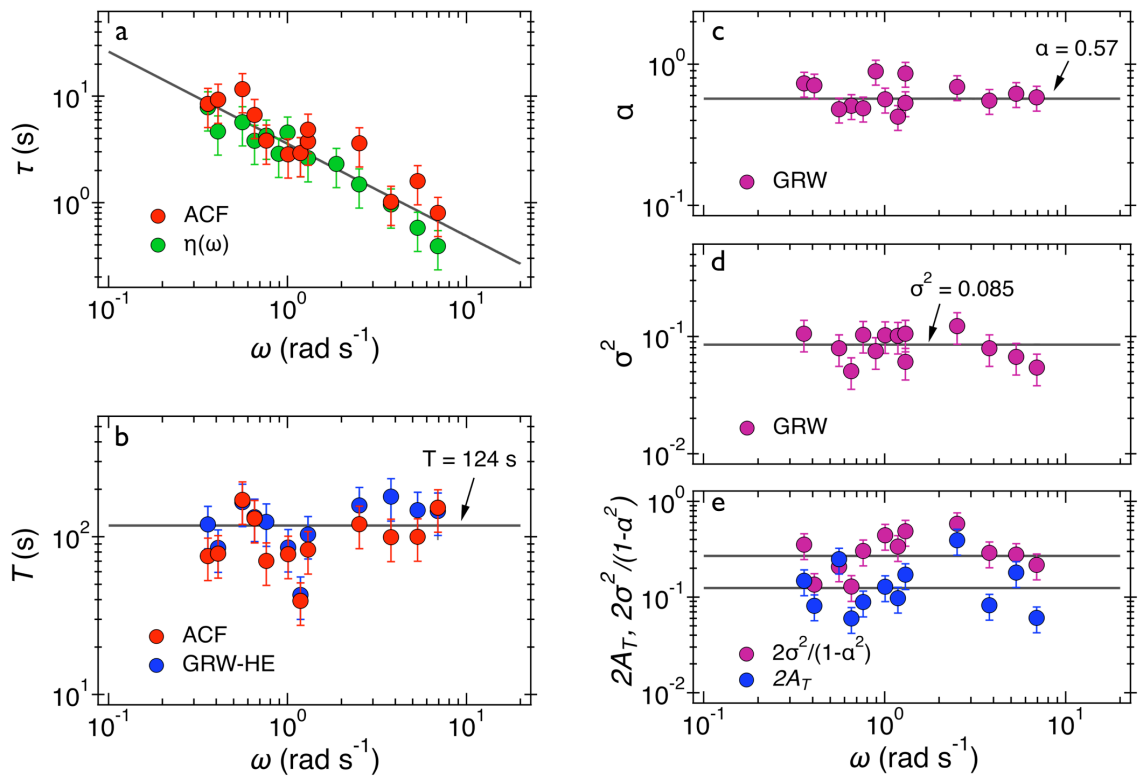

*Figure 7: a) Autocorrelation function decay time τ as a function of the applied angular frequency ω. The straight line is the result of a least square calculations using a power law with exponent -0.87. Also n are the ratios η(ω)/G₀, where the η(ω) is the intracellular shear viscosity (Fig. 4) and G₀ the elastic modulus.[32] b) Oscillation period T obtained from the autocorrelation functions*

and from the Generalized Random Walk model with the Hole effect modification (GRW-HE, Eq. 11). The period $T$ has an average value of 124 s. **c) and d)** Sub-diffusive coefficient $\alpha$ and variance $\sigma^2$ obtained from fitting the variograms with Eq. 6 (GRW model). **e)** Amplitude of the periodic oscillations $A_T$ describing the variogram in the limit of large lag numbers as a function of angular frequency. The adjustment was obtained using the GRW-HE model (Eq. 11). Also displayed is the prefactor $2\sigma^2/(1-\alpha^2)$ featuring in Eq. 6. The figure shows that the oscillations have a smaller amplitude than the GRW saturating values.

| Magnetic wires | $\tau^*$ (s) | $\delta$ | $T$ (s) | $\alpha$ | $\sigma^2$ | $A_T$ |
|---|---|---|---|---|---|---|
| L = 2.8 µm | 3.6 | -0.87 | 115 ± 37 | 0.63 ± 0.16 | 0.15 ± 0.07 | 0.18 ± 0.14 |
| L = 3.0 µm | 3.4 | -0.95 | 131 ± 51 | 0.53 ± 0.13 | 0.12 ± 0.03 | 0.19 ± 0.15 |
| L = 1.9 µm | 9.8 | -1.10 | 145 ± 49 | 0.55 ± 0.09 | 0.12 ± 0.02 | 0.25 ± 0.06 |

**Table I**: Average parameters $\tau^*$, $\delta$, $T$, $\alpha$, $\sigma^2$ and $A_T$ retrieved from fitting ACF and variogram functions using the GRW, GRW-HE, OUP and OUP-HE models. Here $\tau^*$ denotes the ACF relaxation time at $\omega = 1$ rad s$^{-1}$, $\delta$ the exponent of the scaling law $\tau(\omega) \sim \omega^\delta$, $T$ the period of the variogram oscillations (Fig. 5a2-5c2), $\alpha$ the sub-diffusive dynamics coefficient, $\sigma^2$ the variance and $A_T$ the amplitude of the Hole effect (Eq. 11). For each wire studied, the values are averaged over 7 to 14 angular frequencies. The complete set of data can be found in **Supplementary information S8 and S9**.

We now turn to the search of scaling behaviors regarding the previous variogram and cumulative distribution functions. At first, we focus on the initial increase of the variograms (Figs. 5a2, 5b2 and 5c2) over the first seven lags or so, *i.e.* in the range where the Hole effect oscillations contribute marginally to the function and where Eq. 6 applies. The quantity $(1-\alpha^2)\gamma_{GRW}(l)/2\sigma^2$ was thereby evaluated for each time series and plotted as a function of $\alpha^l$ (Fig. 8a). This normalized representation enables to assess the quality of the GRW model adjustments at low lags. The variogram data displayed in Fig. 8a exhibit an excellent superposition at all frequencies and are moreover well described by the function $1-\alpha^l$ shown as the continuous dark grey line (Eq. 6). These outcomes confirm the validity of the GRW model for fitting the intracellular viscosity fluctuations In a second step, we aimed to strengthen the analogy found in the cumulative distribution functions with financial times series, such as the Standard and Poor's 500 stock index and the long-term German bonds (Bund), the data considered being in the time period from September 1991 to August 2001.[33] To this aim, we reexamine the data from Figs. 5a3, 5b3 and 5c3 and establish a master curve by plotting the cdf as a function of the viscosity fluctuations $|\Delta\eta|$ divided by the standard deviation $\sigma(\Delta\eta)$.[33] Subsequently these scaled variations are fitted with a Student's t-distribution function. The results of this approach are shown in Fig. 8b for the three frequencies $\omega$ = 0.65, 1.3 and 5.3 rad s$^{-1}$ already considered. Again, a good superposition of all data is observed, leading to a single Student's t-function (continuous dark grey line in Fig. 8b). The parameter µ, or degrees of freedom, varies between 1 and ∞, depending on the type of probability distribution (1 for a Cauchy distribution and ∞ for a Gaussian). Here µ = 3.5 in line with values obtained by the same analysis for financial markets. The degrees of freedom estimate of the Student's t-distribution for the angular frequencies 0.65, 1.3 and 5.3 rad s-1 are respectively 2.7, 2.6 and 2.1, which compare well with the values reported for the Bund changes.

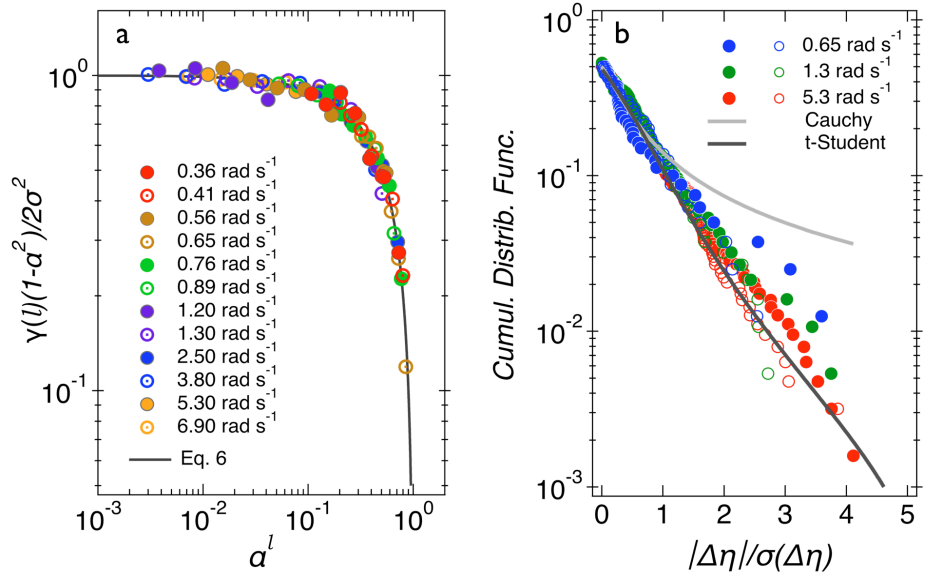

*Figure 8: a) Variogram data derived from intracellular viscosity times series at different frequencies and plotted as $(1 - \alpha^2)\gamma_{GRW}(l)/2\,\sigma^2$ versus $\alpha^l$. The experimental conditions are those of Figs. 1, 2 and 3. The continuous line in dark grey is the function $1 - \alpha^l$ derived from Eq. 6. b) Cumulative distribution function of the viscosity changes for the three frequencies in Fig. 5, corrected for the mean value and standard deviation. Open (resp. closed) symbols are for negative (resp. positive) Δη values. Adjustment using the Cauchy and Student's t-distributions are also indicated. The Student's t-distribution is characterized by mean value, $\langle|\Delta\eta|/\sigma(\Delta\eta)\rangle$ = 0.03, a standard deviation of 0.67 and a degree of freedom, µ = 3.5.*

## 4. Conclusion

The intracellular viscosity of eukaryotic living cells is studied using anisotropic magnetic wires in a configuration that reproduces the shear conditions of rotational rheology. A careful examination of the wire rotation angle *versus* time reveals the existence of anomalous transient responses characterized by intermittent phases of slow and rapid rotation and by large viscosity fluctuations. This stands in contradistinction to non-living matter, where this intermittent behavior is not observed.[38,40,42,57] In the steady rotation phases, the analysis of the wire motion leads to the conclusion that the cytoplasm of murine NIH/3T3 fibroblasts is similar to a viscoelastic liquid and equivalent to a Generalized Maxwell fluid model. The shear viscosity $\eta$ and elastic modulus $G_0$ are retrieved from these measurements and are in agreement with earlier reports.[10,17] In this research, the focus is on temporal fluctuations of the intracellular viscosity observed with fibroblast cells. First, it is found that the fluctuations exhibit large variations around the mean value $\bar{\eta}(t)$ and that the standard variations $\sigma(\Delta\eta)$ are of the order of the mean. For the first time to our knowledge, an approach borrowed from mathematical finance is applied to process statistical data collected from living cells. By treating the time dependent viscosity measurements as a time series, a detailed analysis of the autocorrelation functions and the variograms give evidence of a sub-diffusive mean-reverting process characterized by a frequency independent autoregressive coefficient $\alpha \sim 0.6$ (Eq. 5). This property implies that whatever the amplitude and trend of the fluctuations, the viscosity will drift over time towards its long-term mean $\bar{\eta}(t)$. The second important result, emerging from this analysis, is the evidence of two well separated time scales in the mechanical cell behavior, one around 10 s and associated with the autocorrelation decay time and one around 120 s related to periodic oscillations in the variograms. The correlation decay time shows an unambiguous relation to the viscoelastic properties of the cell, as it is found to fit with the average rheological time $\eta(\omega)/G_0$. The similarity of these characteristic times, obtained from both the time-averaged measurements[32] and from the time-dependent fluctuations, is an indication of the consistency of the current treatment. The biological mechanisms associated with these two time-scales are possibly related to the collective dynamics of the actin network,[27-29] an hypothesis that can be verified using actin stabilizing and depolymerizing drugs.[26] Moreover, the viscosity fluctuations normalized to the standard deviation are described in a satisfactory way by a Student's t-distribution, a model used to quantify the large fluctuations in financial markets.[33] The present outcomes could serve as an illustration of how stochastic data from cell metabolism studies could be analyzed, revealing new features. They also suggest that the proposed method may be applied to other time-dependent measurements of cellular activity, including time series of particle tracking,[54] or optical-tweezers-based microrheology testing the intracellular modulus of Hela cells.[52,58] The technique of micro-rheology and the deeper understanding of time-dependent viscosity can also determine novel biomarker candidates for studying the effect of pharmaceutical and physical therapies on cellular metabolism.

## Acknowledgments

ANR (Agence Nationale de la Recherche) and CGI (Commissariat à l'Investissement d'Avenir) are gratefully acknowledged for their financial support of this work through Labex SEAM (Science and Engineering for Advanced Materials and devices) ANR 11 LABX 086, ANR 11 IDEX 05 02. We acknowledge the ImagoSeine facility (Jacques Monod Institute, Paris, France), and the France BioImaging infrastructure supported by the French National Research Agency (ANR-10-INSB-04, « Investments for the future »). This research was supported in part by the Agence Nationale de la Recherche under the contract ANR-13-BS08-0015 (PANORAMA), ANR-12-CHEX-0011 (PULMONANO), ANR-15-CE18-0024-01 (ICONS), ANR-17-CE09-0017 (AlveolusMimics) and by Solvay.

## Supplementary Information

S1: Geometric characteristics of the wires used in this work; S2: Spontaneous internalization of magnetic wires in fibroblasts; S3: Complementary data showing the critical frequency $\omega_C(t)$ and the shear viscosity $\eta(t)$ time series; S4: Details on the method used to determine the time depend-ent viscosity and removal of outliers; S5: Parameters derived from the Generalized Random Walk and Ornstein-Ühlenbeck Process (OUP) models; S6: Analysis of shear elastic modulus fluctuations obtained by means of optical tweezers in HELA cells (PhD thesis from Ming-Tzo Wei, Lehigh University, 2014); S7: Adjustment of the cumulative distribution function (cdf) with Gaussian fits; S8: Frequency dependent autocorrelation time $\tau(\omega)$ obtained from different models and from 3 different wires; S9: Values of the parameters obtained by fitting data series obtained on three different wires

## TOC Image

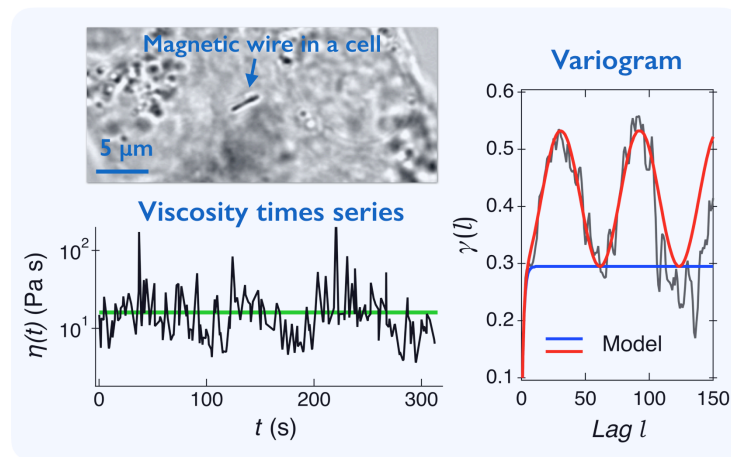